----------------------------------------------------------------------------
\documentstyle[eqsecnum,aps]{revtex}
\begin{document}

\title{U(2) algebraic model \\
applied to stretching vibrational spectra of tetrahedral molecules }

\author{Xi-Wen Hou$^{1,2)}$ and Zhong-Qi Ma$^{1)}$} 

\address{~$^{1)}$~Institute of High Energy Physics, 
P.O. Box 918(4), Beijing 100039, The People's Republic of China \\
~$^{2)}$~Department of Physics, University of Three Gorges, 
Yichang 443000, The People's Republic of China }

\maketitle

\vspace{50mm}
\begin{abstract}
The highly excited stretching vibrational energy levels and the 
intensities of infrared transitions in tetrahedral molecules are 
studied in a U(2) algebraic model. Its applications to silane and
silicon tetrafluoride are presented with smaller standard deviations
than those of other models.
\end{abstract}

\vspace{20mm}

\newpage

\section{INTRODUCTION}

In recent years, algebraic models, such as Lie algebraic methods
(Iachello and Levine, 1995; Bijker $et$ $al$., 1995) and 
the boson-realization model (Ma $et$ $al.$, 1996), have been proposed
for descriptions of vibrations, rotations, 
and rotation-vibration interactions in polyatomic molecules. 
Lie algebraic methods for diatomic molecules have been modified
by the corresponding quantum algebra (Alvarez $et$ $al.$, 1994; Chang, 1995), 
and the boson-realization model have been developed for the 
higher vibrational states of polyatomic molecules in terms of
q-deformed oscillators (Xie $et$ $al.$, 1996; Hou $et$ $al.$, 1997).

In Lie algebraic approaches, U(4) and U(2) algebraic model have been 
extensively used. The U(4) model took the rotation and the vibration
into account simultaneously but became quite complicated when the number
of atoms in a molecule increased to larger than four.
The U(2) model was particularly successful
in explaining stretching vibrations of polyatomic molecules such as
benzene-like and octahedral systems (Iachello and Oss, 1991; Chen 
$et$ $al.$, 1996). This model was extended to
deal with both stretching and bending vibrations in triatomic molecules
(Frank $et$ $al.$, 1996). Recently, a U(5) algebraic model was 
introduced for higher excited stretching modes and infrared intensities 
of tetrahedral molecules (Leroy $et$ $al.$, 1996; Leroy and Boujut, 1997). 
However, the U(5) model was less feasible
than the U(2) model when the bending vibrations were considered.
  
In this paper, we will use the U(2) algebraic model to study the stretching 
vibrations and intensities of infrared transition of silane SiH$_4$ and 
silicon tetrafluoride SiF$_4$. Our results are quite good in comparison
with those of the U(5) algebraic model for SiH$_4$ (Leroy $et$ $al.$, 1996) 
and the local mode model (Della Valle, 1988). Results for stretching and 
bending vibrations in the U(2) model and the boson-realization model
will be presented in a subsequent publication.

\section{U(2) ALGEBRAIC MODEL}

For a tetrahedral molecule XY$_4$, we introduce four U(2)
algebras to describe the vibrations of four X-Y bonds.
The molecular dynamical group is
$$ U_1(2) \otimes U_2(2) \otimes U_3(2) \otimes U_4(2),$$

\noindent
where each U$_{i}$(2) ($1\leq i \leq 4$) is
generated by the operators $\{ \hat{N}_{i},\hat{J}_{+,i} 
\hat{J}_{-,i}, \hat{J}_{0,i} \}$, 
satisfying the following commutation relations (Frank $et$ $al.$, 1996):  
$$\begin{array}{ll}
~[\hat{J}_{0,i}, \hat{J}_{\pm,j}]=\pm \delta_{ij}
\hat{J}_{\pm,i},~~~
&[\hat{J}_{+,i}, \hat{J}_{-,j}]=2 \delta_{ij} 
\hat{J}_{0,i},\\
~[\hat{N}_{i}, \hat{J}_{0,j}]=0,  
~~&[\hat{N}_{i}, \hat{J}_{\pm,j}]=0 .
\end{array}  $$

\noindent
where $\hat{N}_{i}$ is related with the Casimir operator
of U(2):
$$2\hat{J}_{0,i}^{2}+\hat{J}_{+,i}\hat{J}_{-,i}
+\hat{J}_{-,i}\hat{J}_{+,i}
=\hat{N}_{i}(\hat{N}_{i}/2+1). $$

\noindent
Denote by $v_{i}$ the number of quanta in the $i$th
bond. The local basis states for each bond are labeled by 
the eigenvalue $N_{i}$ of $\hat{N}_{i}$
and $v_{i}$, and written as $|N_{i},v_{i}\rangle$. 
Their products provide the local bases: 
$$|N_1,v_1\rangle  |N_2,v_2\rangle |N_3,v_3\rangle |N_{4},v_{4}\rangle 
 \equiv  |N_{i},v_{i}\rangle .   $$

\noindent
where those $N_{i}$ are equal to each other, 
$N_{i}= N$, due to equivalent bonds .

There are three kinds of O(2) invariant combinations
of those generators: 
$$\begin{array}{rl}
\hat{H}_{i}&=~(\hat{J}_{+,i}\hat{J}_{-,i}  
+\hat{J}_{-,i}, \hat{J}_{+,i})/2~-~\hat{N}_{i}/2, \\
\hat{H}_{ij}&=~2\hat{J}_{0,i}\hat{J}_{0,j}
~-~ \hat{N}_{i}\hat{N}_{j}/2,~~~ i \neq j,  \\
\hat{V}_{ij}&=\hat{J}_{+,i}\hat{J}_{-,j}~+~  
\hat{J}_{-,i}\hat{J}_{+,j}, ~~~ i \neq j.
\end{array} \eqno (1) $$

\noindent
Their matrix elements in the local bases are given by Frank $et$ $al.$
(1996). The operator $\hat{H}_{i}$ corresponds to the energy of the
$i$th Morse oscillator. The operators $\hat{H}_{ij}$
describe the anharmonic terms with the type $v_{i}v_{j}$, while the 
operators $\hat{V}_{ij}$ describe the interbond couplings which, in 
configuration space, are of the type ${\bf r}_{i}{\bf r}_{j}$, 
where ${\bf r}_{i}$ and ${\bf r}_{j}$ are the displacement
vectors of bonds $i$ and $j$ from their equilibrium positions.
 
The Hamiltonian, if restricted at the quadratic terms, is
expressed in terms of those three kinds of operators as follows:
$$\begin{array}{rl}
H&~=~\displaystyle \lambda_{1}~ \sum_{i=1}~ \hat{H}_i
~+~\lambda_{2}~\displaystyle \sum_{i\neq j}~ \hat{H}_{ij}
~+~\lambda_{3}~\displaystyle \sum_{i\neq j}~ \hat{V}_{ij}  
~+~\lambda_{4}~\displaystyle \sum_{i=1}~ (\hat{H}_i)^2 \\
&~~~+~\lambda_{5}~\displaystyle \sum_{i,j\neq k}~ \hat{H}_{i} \hat{H}_{jk}
~+~\lambda_{6}~\displaystyle \sum_{i\neq j}~ \hat{H}_i \hat{H}_j
~+~\lambda_{7}~\displaystyle \sum_{i\neq j}~ \hat{H}_{i} \hat{H}_{ij}
~+~\lambda_{8}~\displaystyle \sum_{i\neq j}~ \hat{H}_{i} \hat{V}_{ij} \\
&~~~+~\lambda_{9}~\displaystyle \sum_{i,j\neq k}~ \hat{H}_{i} \hat{V}_{jk} 
~+~\lambda_{10}~\displaystyle \sum_{i\neq j}~ (\hat{H}_{ij})^2
~+~\lambda_{11}~\displaystyle \sum_{i\neq j\neq k}~ \hat{H}_{ij}\hat{H}_{ik}
~+~\lambda_{12}~\displaystyle \sum_{i\neq j,k\neq l}~ \hat{H}_{ij}\hat{H}_{kl}\\
&~~~+~\lambda_{13}~\displaystyle \sum_{i\neq j}~ (\hat{V}_{ij})^2  
~+~\lambda_{14}~\displaystyle \sum_{i\neq j\neq k}~ \hat{V}_{ij}\hat{V}_{ik} 
~+~\lambda_{15}~\displaystyle \sum_{i\neq j,k\neq l}~ \hat{V}_{ij}\hat{V}_{kl}\\
&~~~+~\lambda_{16}~\displaystyle \sum_{i\neq j}~ \hat{H}_{ij}\hat{V}_{ij} 
~+~\lambda_{17}~\displaystyle \sum_{i\neq j\neq k}~ \hat{H}_{ij}\hat{V}_{ik}
~+~\lambda_{18}~\displaystyle \sum_{i\neq j,k\neq l}~ \hat{H}_{ij}\hat{V}_{kl} ,
\end{array}   \eqno (2) $$

\noindent
where all $\lambda$'s are the coupling parameters.
The Hamiltonian preserves the quantum number $V=\sum v_{i}$.

We now apply this model to study the stretching vibrational spectra 
of SiH$_4$ and SiF$_4$. The calculation for 
energy levels has been greatly simplified since the symmetrized bases 
are used. The boson number $N$ is taken 
to be 60 for SiH$_4$, and 100 for SiF$_4$.

\vspace{3mm}
\begin{center}

\fbox{Table I}

\vspace{2mm}

\fbox{Table II}

\end{center}

\vspace{3mm}

For SiH$_4$, we choose five parameters, the same as the number 
of parameters used by Leroy $et$ $al.$ (1996), to calculate 
the vibrational levels. It was found that these five parameters 
$\lambda_{j}$, $1 \leq j \leq 5$, can give better results.
That means to set $\lambda_i$=0, $6 \leq i \leq 18$. Fitting the 
observed data for SiH$_4$ from the compilation of Leroy $et$ $al.$
(1996), we obtain the parameter values and the calculated energy levels,
listed in Table I.  For SiF$_4$, the better results can be obtained
in terms of only three parameters $\lambda_{1}$, $\lambda_{2}$, and 
$\lambda_{3}$. Its observed data from McDowell $et$ $al.$ (1982),
calculated values, and corresponding parameters are also listed in
Table I. For comparison, the levels for the
two molecules calculated by the local model model (LMM) 
(Della Valle, 1988) and the recently calculated results for SiH$_4$ 
by the U(5) model (U(5)M) (Leroy $et$ $al.$, 1996) are 
given in Table I, together with their standard deviations (SD). The 
calculated vibrational energy levels given in Table I are the 
differences between the observed values and the calculated ones. 

\section{INTENSITIES OF INFRARED TRANSITION}

In the following we will introduce infrared transition operator.
The calculated intensities can be used to check assignments and to
study the intramolecular energy relaxation in tetrahedral molecules.

For the considered systems, the infrared active mode is $F_2$. The 
absolute absorption intensities from a state $v'$ to $v$ are given by
$$\begin{array}{rl}
I_{vv'}&=~\nu_{vv'}P_{vv'} ,  \\
P_{vv'}&=~|\langle v|\hat{T}_x|v'\rangle|^{2}+
|\langle v|\hat{T}_y|v'\rangle |^{2}+
|\langle v|\hat{T}_z|v'\rangle |^{2},
\end{array}  \eqno (3) $$

\noindent
where $\nu_{vv'}$ is the frequency of the observed transition,
$\hat{T}_x$, $\hat{T}_y$, and $\hat{T}_z$ correspond to the three
components of the infrared transition operator $\hat{T}$, and
the state $|v\rangle$ denotes $|N_{i}, v_{i}\rangle$ 
for short. All other constants are absorbed in the normalization 
of the operator $\hat{T}$. The three components of $\hat{T}$ are
$$\begin{array}{rl}
\hat{T}_x&=~\alpha~(\hat{t}_1-\hat{t}_2+\hat{t}_3-\hat{t}_4),\\
\hat{T}_y&=~\alpha~(\hat{t}_1-\hat{t}_2-\hat{t}_3+\hat{t}_4), \\
\hat{T}_z&=~\alpha~(\hat{t}_1+\hat{t}_2-\hat{t}_3-\hat{t}_4),
\end{array}  \eqno (4) $$

\noindent
where $\alpha$ is the parameter. The matrix elements of $\hat{t}_{i}$ are 
taken to be (Iachello and Oss, 1993)
$$\langle \hat{N}_{i},v_{i}|\hat{t}_{i}
|\hat{N}_{i},v'_{i}\rangle~=~exp(-\beta_{i}
|v_{i}-v'_{i}|). \eqno (5) $$

\noindent
Those $\beta_{i}$ for equivalent bonds are equal to each other,
and denoted by a common symbol $\beta$. 

The two parameters in the transition operator of (4) and (5)
will be determined by fitting observed data. 
Due to the Wigner-Eckart theorem, it is sufficient to calculate
only the $z$ component of the transition operator, $\hat{T_z}$,
in the symmetrized bases. In Table II we only list part of calculated 
intensities to compare with known observed data, from that the two 
parameters are determined. The standard deviation in our fitting 
is 1.265 for SiH$_4$, and 2.512 for SiF$_4$, while that was 
1.415 for SiH$_4$ by Leroy $et$ $al.$ (1996).

\section{CONCLUSION}

We have used a U(2) algebraic model for studying the
stretching vibrations and infrared intensities of a
tetrahedral molecule. The model Hamiltonian and the model transition 
operator have provided better fits to the published experimental 
data of silane and silicon tetrafluoride than the local mode model 
(Della Valle, 1988) and the U(5) algebraic approach (Leroy $et$ $al.$, 1996) 
to silane. In addition, U(2) algebraic model
can be applied to the stretching and bending vibrations of 
other medium-size and large molecules. Furthermore, this model
can be studied by the corresponding quantum algebra (Bonatsos and 
Daskaloyannis, 1993). Investigations on those subjects are under way.

\acknowledgments 
This work was supported by 
the National Natural Science Foundation of China and Grant No. 
LWTZ-1298 of the Chinese Academy of Sciences.

\vspace{5mm}
{\bf REFERENCES}

\vspace{5mm}

\noindent
Alvarez, R. N., Bonatsos, D., and Smirnov, Y. F., (1994) Phys. Rev. A
{\bf 50}, 1088. \\
Bernheim, R. A., Lampe, F. W., O'Keefe, J. F., and Qualey III, J. R.,
(1984), J. Chem. Phys. {\bf 80}, 5906. \\
Bijker, R., Dieperink, A. E. L., and Leviatan, A., (1995) Phys. Rev. A
{\bf 52}, 2786. \\
Bonatsos, D. and Daskaloyannis, C., (1993), Phys. Rev. A {\bf 48},
3611. \\
Chang, Z., (1995), Phys. Reports. {\bf 262}, 137. \\
Chen, J. Q., Iachello, F. and Ping, J. L., (1996), J. Chem. Phys.
{\bf 104}, 815. \\
Della Valle, R. G., (1988), Mol. Phys. {\bf 63}, 611. \\
Frank, A., Lemus, R., Bijker, R., P\'erez-Bernal, F., and  Arias, J. M.,
(1996), Ann. of Phys. {\bf 252}, 211. \\
Hou, X. W., Xie, M., and Ma, Z. Q., (1997), Phys. Rev. A., {\bf 55}, 3401;
Int. J. Theor. Phys. {\bf 36}, 1153. \\
Iachello, F. and Levine, R. D., (1995), Algebraic Theory of Molecules
(Oxford University, Oxford).  \\
Iachello, F. and Oss, S., (1993), J. Chem. Phys. {\bf 99}, 7337. \\
Iachello, F. and Oss, S., (1991), Phys. Rev. Lett. {\bf 66}, 2976. \\
Leroy, C., Collin, F., and Lo\"ete, M., (1996), J. Mol. Spectrosc.
{\bf 175}, 289.\\
Leroy, C. and Boujut, V., (1997), J. Mol. Spectrosc. {\bf 181}, 
127. \\
Ma, Z. Q., Hou, X. W., and Xie, M., (1996), Phys. Rev. A {\bf 53},
2173. \\
McDowell, R. S., Reisfeld, M. J., Patterson, C. W., 
Krohn, B, J., Vasquez, M. C., and Laguna, G. A., (1982),
J. Chem. Phys. {\bf 77}, 4337. \\
Xie, M., Hou, X. W., and Ma, Z. Q., (1996), Chem. Phys. Lett. {\bf 262}, 1.\\

\newpage
\begin{center}
{\small Table I. Observed and calculated energy levels, and 
parameters (in cm$^{-1}$)}

\vspace{3mm}
\begin{tabular}{||cccc|ccc||}
\hline
\hline
  & SiH$_4$ & & & & SiF$_4$ &  \\
~E$_{obs}$ & LMM & U(5)M & U(2)M  & ~E$_{obs}$ & LMM &
U(2)M  \\ 
 \hline
2186.8730 & 0.5339 & 0.7378 & 0.4476 & 800.6 & -5.1243 & -1.5226 \\
2189.1901 & -0.0652& -0.5527& -0.0765& 1031.3968& -3.1250&1.7942 \\
4374.5600 & 0.3247 & -0.1007& -0.0503& 2059.1 & -0.2942 & 4.4876 \\
4308.3800 & -1.4258& -0.8963& -0.6650& 1828.17& -1.5378 & -0.0686\\
4380.2800 & 1.7690 & 0.2231 & -0.4878& 2623.8 & 4.2976 & 0.9521\\
4378.4000 & 1.3154 & 0.1421 & -0.6465& 3068.5 & 0.5720 & -4.1160\\
\cline{5-7}
4309.3485 & -0.5244& 0.0617 & 0.2095 & SD     & 3.090  & 2.066 \\
\cline{5-7}
6496.1300 & 0.7661 & -0.0028& 0.0190 &  &     &      \\
6361.9800 & -2.1182& 0.1450 & 0.5381 &  &     &      \\  
6500.3000 & 2.5976 & 1.4384 & 1.3330 &  &     &    \\   \cline{5-7}
6502.8800 & 2.2612 & 0.4108 & 0.4409 & Param. & SiH$_4$  & SiF$_4$ \\ 
\cline{5-7}
6362.0800 & -2.0186& 0.2448 & 0.6372 & $\lambda_1$ &30.2883&1.1974 \\ 
6497.4810 & 0.9458 & -0.4104& -0.0605& $\lambda_2$ &-2.2592&-2.8473\\
6500.6000 & 1.0278 & -0.1297& -0.05781& $\lambda_3$ &-0.0147&-0.5687\\
8347.4000 & -3.8164& 0.1031 & 0.2949 & $\lambda_4$ &-0.0021&   \\
10267.2200& -3.9023& 0.4100 & 0.1748 & $\lambda_5$ &-0.0003&   \\  
\cline{5-7}
12121.2000& -2.7061& -0.6399& -0.1973&    &   &  \\
13914.4000& 4.8506 & 0.2227 & 0.3037 &    &   &   \\ \cline{1-4}
SD     &  2.465  & 0.610 & 0.573     & & & \\
\hline
\hline
\end{tabular}
\end{center}

\vspace{10mm}
\begin{center}
{ TABLE II. Observed and Calculated Relative Intensities}
\end{center}

\begin{center}
\begin{tabular}{||c|ccccc|ccccc||}
\hline
\hline
 & & & SiH$_4$ & & & & & SiF$_4$ & & \\
 & $\alpha$ & 8.130 & & $\beta$& 0.866 & $\alpha$ & 37.977& & $\beta$
& 3.540 \\ 
\hline
E$_{calc}$& 10267.04 & 12121.40& 13914.10 & 15645.57 & 17318.60 & 
1029.60 & 1828.24 & 2054.60 & 2622.84 & 3072.62  \\
Obs.$^{a}$ &    & 100 & 21 & 2.4 & 0.6 & 5000 & 7 & 1.2 & 0.015 & 0.015 \\
Calc.$^{b}$& 530 & 100 & 20.18 & 4.2 & 0.9 &  & & & &\\
Calc. & 479 & 100 & 20.33 & 4.04 & 0.79 & 5000 & 3.79 & 4.14 &
0.005 & 0.003 \\
\hline
\hline
\end{tabular}
\end{center}
{\footnotesize ~$^{a}$Observed data for SiH$_4$ from Bernheim $et$ $al.$ (1984),
 and for SiF$_4$ from McDowell $et$ $al.$ (1982). 
$^{b}$Calculated by Leroy $et$ $al.$ (1996).}

\vspace{5mm}

\end{document}